\documentclass[journal]{IEEEtran}
\usepackage{cite}
\usepackage{color}
\usepackage{times}
\usepackage{soul}
\usepackage{url}
\usepackage[hidelinks]{hyperref}
\usepackage[utf8]{inputenc}
\usepackage[small]{caption}
\usepackage{graphicx}
\usepackage{amsmath}
\usepackage{amsthm}
\usepackage{booktabs,bigstrut}
\usepackage{algorithm}
\usepackage{algorithmic}
\graphicspath{{Figure/}}
\usepackage{multirow}
\usepackage{multicol}
\usepackage[american]{babel}
\usepackage{microtype}

\begin{document}
	
	\title{When Image Decomposition Meets Deep Learning: A Novel Infrared and Visible Image Fusion Method}
	
	\author{Zixiang Zhao, 
		Jiangshe Zhang,
		Shuang Xu, 
		Kai Sun,
		Chunxia Zhang, 
		Junmin Liu,~\IEEEmembership{Member,~IEEE}	
		\thanks{This manuscript is submitted to IEEE TNNLS Special Issue on ``Effective Feature Fusion in Deep Neural Networks''.}
		\thanks{Corresponding author: Jiangshe Zhang, E-mail: jszhang@mail.xjtu.edu.cn.}
		\thanks{Z.X. Zhao, S.Xu, K. Sun, C.X. Zhang, J.M Liu and J.S. Zhang are with the School of Mathematics and Statistics, Xi’an Jiaotong University, Xi’an, Shaanxi, 710049, P.R.China.}
		\thanks{This article was presented in part at the IJCAI 2020 \cite{zhaoijcai2020}.}
		\thanks{The research is supported by the National Key Research and Development Program of China under grant 2018AAA0102201, the National Natural Science Foundation of China under grant 61976174, 11671317 and 61877049, the Fundamental Research Funds for the Central Universities under grant number xzy022019059.}
	}
	
	\markboth{IEEE Transactions on Neural Networks and Learning Systems,~Vol.~X, No.~X, XXX.~2020}%
	{Z.X. Zhao \MakeLowercase{\textit{et al.}}: When Image Decomposition Meets Deep Learning: A Novel Infrared and Visible Image Fusion Method}
	
	\maketitle
	
	\begin{abstract}
		Infrared and visible image fusion, as a hot topic in image processing and image enhancement, aims to produce fused images retaining the detail texture information in visible images and the thermal radiation information in infrared images. 
		A critical step for this issue is to decompose features in different scales and to merge them separately.
		In this paper, we propose a novel dual-stream auto-encoder~(AE) based fusion network. The core idea is that the encoder decomposes an image into base and detail feature maps with low- and high-frequency information, respectively, and that the decoder is responsible for the original image reconstruction. 
		To this end, a well-designed loss function is established to make the base/detail feature maps similar/dissimilar. 
		In the test phase, base and detail feature maps are respectively merged via an additional fusion layer, which contains a saliency weighted-based spatial attention module and a channel attention module to adaptively preserve more information from source images and to highlight the objects. Then the fused image is recovered by the decoder. 
		Qualitative and quantitative results demonstrate that our method can generate fusion images containing highlighted targets and abundant detail texture information with strong reproducibility and meanwhile is superior to the state-of-the-art (SOTA) approaches.
	\end{abstract}
	
	
	\IEEEpeerreviewmaketitle

	\section{Introduction}\label{sec:1}
	\IEEEPARstart{I}{mage} fusion, whose principle is to learn the complementary and comprehensive information from source images acquired by different sensors for the same scene~\cite{meher2019a}, has been an important image processing technique for image enhancement and information fusion. 
	Broadly speaking, according to distinct application environments, it can be roughly divided into three categories, digital image fusion (multi-exposure~\cite{DBLP:journals/tip/MaDZFW20,DBLP:journals/tip/MaLYWMZ17} and multi-focus fusion~\cite{DBLP:journals/tmm/XiaoOTBL20,DBLP:journals/tmm/GuoNCZMH19}), multi-modality image fusion (infrared/visible fusion~\cite{ma2019infrared,li2018densefuse} and medical image fusion~\cite{DBLP:journals/tmm/BhatnagarWL13,DBLP:journals/tmm/BernalYLKMRB18}) and remote sensing image fusion (multi-spectral~(MS)/panchromatic~(PAN) fusion~\cite{DBLP:journals/tgrs/00100YJ19,DBLP:journals/tgrs/LiL09} and MS/hyper-spectral~(HS) fusion~\cite{XuALZZL20,9069930}).
	
	Infrared and visible image fusion, abbreviated as IVIF, aims at blending the thermal radiation information in the infrared images and the detailed texture information in the visible images. It has been proved that IVIF benefits to many issues, including surveillance~\cite{bhatnagar2015novel}, modern military and fire rescue tasks~\cite{lahoud2018ar,hu2017adaptive}, face recognition~\cite{ma2016infrared}, etc.
	As is well-known, infrared light has strong penetrating power, so infrared images containing the thermal radiation information are robust to illumination changes and obstacles. However, the side effect is that, infrared images are often with low spatial resolution and poor texture detail information.
	On the contrary, visible images have high spatial resolution along with abundant texture and gradient information, but they are susceptible to illumination alteration, light reflection and obstructions. 
	Therefore, infrared and visible images will be potentially conducive to target recognition and object tracking.
	
	Generally, the IVIF algorithms can be separated into two categories: traditional methods\footnote{Traditional methods, distinguished from deep learning methods, mainly refer to methods that do not use deep learning.} and deep learning methods. Specifically, representative traditional methods include image multi-scale transformation, sparse representation, subspace learning, the saliency degree and Bayesian-based methods.
	These methods mainly process the source images from different perspectives, based on corresponding prior knowledge.
	Methods in multi-scale transformation group~\cite{liu2018deep,li2011performance,pajares2004wavelet,zhang1999categorization} decompose the input image into multiple layers with different kinds of features, and then design specific fusion algorithms for each layer.
	The sparse representation methods~\cite{yang2014visual,wang2014fusion, li2012group} are based on the image sparse prior, i.e., the image can be represented by a linear combination of over-complete dictionary with sparse coefficients. Therefore, the image fusion task can be transformed into a fusion of the coefficients.
	For the subspace group~\cite{bavirisetti2017multi,kong2014adaptive,FU2016114}, dimension reduction operations such as principal component analysis are performed on the input image to extract low-dimensional intrinsic features.
	The saliency methods~\cite{bavirisetti2016two,zhang2017infrared,zhao2014infrared} separate the target and the background parts of the original image by the different salient degrees of the foreground target and background, then diverse fusion algorithms for the corresponding parts are designed respectively.
	After formulating the fusion model into a regression issue, the Bayesian-based method~\cite{ZHAO2020107734} casts the optimization problem into a statistical inference issue for latent variable parameters. As a result, the pixel-wise fusion weights can be adaptive to the source images.
	
	With the rapid development of computer vision, deep learning (DL) has become an efficient tool in IVIF area. DL-based fusion algorithms can be divided into four basic groups.
	The first group is Generative Adversarial Networks (GANs) based methods. In these works~\cite{ma2019fusiongan,ma2020infrared}, the image fusion task is formulated as an adversarial game, in which the generator creates a fusion image containing the radiation and texture information of the source images, and the discriminator adds more details to the generated fusion image. These end-to-end models can avoid manually designing fusion rules.
	The second group is the unified model category~\cite{9151265} which can be applied to multiple fusion tasks. By using different types of fusion datasets to complete continual learning, the model performance is promoted mutually and the issues of unsupervised learning for fusion, e.g., the demand of ground-truth and specifically designed evaluation metrics, can be solved effectively.
	The third is pre-trained convolutional neural network~(CNN) group~\cite{li2018infrared,lahoud2019fast}. As an extension of image multi-scale transformation, methods in this group transform images from the spatial domain to base and detail domains by means of filters or optimization-based methods. Base images are simply averaged. Since there are high-frequency textures in detail images, these methods fuse feature maps of detail images extracted from a pre-trained neural network (for example, VGG-19 \cite{simonyan2014very}). At last, a fusion image is recovered by merging the fused base and detail images. 
	The fourth group consists of various AE-based methods \cite{li2018densefuse,DBLP:journals/tim/LiWD20}. In the training phase, an AE network is trained to extract features from source images. In the test phase, feature maps are merged respectively, which then pass through the decoder to reconstruct a fusion image. In summary, deep neural networks (DNNs) are often employed to extract features of input images and then a certain fusion strategy is exploited to merge features and complete the image fusion task.
	
	One shortcoming of the pre-trained CNN group is worth pointing out, i.e., DL is used only in the fusion stage, and filters or optimization-based methods are employed in the decomposition stage, which leads to a rough decomposition effect. To overcome this shortcoming, by combining principles of the second and the third groups in DL-based methods, we propose a novel dual-stream IVIF network, called deep image decomposition based IVIF (DIDFuse), to exploiting DL in both feature decomposition and image fusion.
 	We first train a dual-stream AE network that can be used for feature separation. The encoder and the decoder are responsible for image decomposition and reconstruction, respectively. In training, we decompose the source images into base and detail feature maps containing low- and high-frequency information with large- and small-region pixel intensity changes by a novel loss function. While in testing, base and detail feature maps of test pairs are separately fused according to an extra information-retaining fusion layer based on a saliency weighted-based spatial attention module and a channel attention module. Then the fused image can be acquired through the decoder. 
	Our contributions are summarized into three-fold:
	
	(1) To the best of our knowledge, this is the first deep image decomposition model for IVIF task, where both decomposition and fusion are accomplished via a deep AE network. In the training stage, the feature decomposition component of loss function forces base and detail feature maps of two source images similar/dissimilar to accomplish the two-scale decomposition and separate base/detail features effectively. Simultaneously, the image reconstruction component of the loss function maintains pixel intensities between source and reconstructed images, as well as gradient details of the visible image. In result, the well-designed loss function and the novel AE architecture accurately separate the multi-scale complementary information, while retaining the highlight information of objects and the detailed texture information in entire images.
	
	(2) We propose an adaptive information-retaining fusion layer via a saliency-weighted spatial attention module and a channel attention module to preserve significant features useful for fusion. 
	The employment of spatial and channel attention mechanism allows the fusion layer to better retain the information of interesting objects and salient features.
	Experiments imply that the effect of our proposed fusion layer is better than the traditional weighted-average addition and $\ell_1$-norm addition, etc.
	
	(3) As far as we know, the performance of existing IVIF methods \cite{ma2016infrared,li2018densefuse,zhang2017infrared,li2018infrared} is only verified on a limited number of hand-picked examples in TNO dataset. Their results may not be as superior as described in their papers if the testset contains various scenery. To measure the fusion effectiveness of our model more convincingly, we enlarge the number of test datasets to five, including TNO, FLIR and three scenery in NIR. In total, there are 236 test images with indoor/outdoor scenes and daylight/nightlight illuminations. Compared with the SOTA methods, our method can robustly create fusion images with brighter targets and richer details.
	
	The previous version of this work was published in \cite{zhaoijcai2020}. Compared with it, we have made the following improvements: 
	First, the adaptive information-retaining fusion layer based on spatial/channel attention mechanism is exploited to replace the traditional ``summation'' strategy in the original paper and achieve adaptive retention of significant information in source images.
	Second, various ablation experiments are supplemented to verify the effectiveness of different modules in our network. 
	Third, we add three more test scenery in NIR fusion dataset to further prove the effectiveness of our model for different objects and scenery. In addition, two state-of-the-art (SOTA) methods~\cite{9151265,DBLP:journals/tim/LiWD20} published during the conference review period are added in the qualitative and quantitative comparison.
	Fourth, more details and analysis for the experiments are provided, such as the calculation of evaluation metrics, detailed analysis for qualitative results, and determination of hyperparameters.
	
	The remaining article is arranged as follows. A brief introduction of related work is exhibited in section~\ref{sec:2}. The mechanism and architecture of our proposed network are elaborated in section~\ref{sec:3}. Then, experimental results are shown in section~\ref{sec:4}. At last, some conclusions and the future plan are drawn in section~\ref{sec:5}.
	
	\section{Related Work}\label{sec:2}
	Since our network structure is closely related to U-Net, we introduce U-Net architecture in section \ref{sec:2_1}. Then, traditional two-scale image decomposition methods are briefly reviewed in section \ref{ts_de}. Lastly, we give a brief introduction about deep learning applied in IVIF task in section \ref{sec:2_3}.
	
	\subsection{U-Net and Skip Connection}\label{sec:2_1}
	U-Net~\cite{ronneberger2015u}, a famous network architecture in biomedical image segmentation, consists of a contracting path for feature extraction and an expanding path for precise localization. Compared with AE, there is a channel-wise concatenation of corresponding feature maps from contracting and expanding paths in U-Net. In this manner, it can extract “thicker” features that help preserve image texture details during downsampling. In literature \cite{mao2016image}, a U-Net-like symmetric network is used for image restoration. It employs skip connection technique, where feature maps of convolution layers are added to corresponding deconvolution layers to enhance the information extraction capability of the neural network and to accelerate convergence.
	\subsection{Two-Scale Decomposition}\label{ts_de}
	As a subset of multi-scale transformation, two-scale decomposition in IVIF decomposes an original image into base and detail images with low- and high-frequency information, respectively.
	In \cite{li2018densefuse}, given an image $I$, they obtained the base image $I^{b}$ by solving the following optimization problem,
	$$I^{b} = \arg\min ||I-I^{ b}||_F^2+\lambda (||g_x*I^{ b}||_F^2+||g_y*I^{ b}||_F^2),$$
	where $*$ denotes a convolution operator, and $g_x=[-1,1]$ and $g_y=[-1,1]^T$ are gradient operators. Then, the detail image is acquired by $I^{d}=I-I^{b}$. Similarly, a box filter is used to get the base image in \cite{lahoud2019fast}, and the method of obtaining the detail image is the same as that of \cite{li2018densefuse}. After decomposition, base and detail images are separately fused with different criteria. At last, the fused image is reconstructed by combining fused base and detail images.
	\subsection{Deep Learning in IVIF Fusion}\label{sec:2_3}
	Deep learning, as a black-box feature extraction tool, is often used in various IVIF methods. In section~\ref{sec:1}, we divide the DL-based methods into four categories: GAN-based group, unified model group, pre-training model group and AE-based group.
	For the GAN-based group, in FusionGAN \cite{ma2019fusiongan}, a generator creates fused images with infrared thermal radiation and visible gradient information, a discriminator forces the fused results to have more details from the visible images. In the light of Conditional GANs \cite{mirza2014conditional}, detail preserving GAN \cite{ma2020infrared} changes the loss function of FusionGAN for improving the quality of detail information and sharpening the target boundary.
	For the unified model group, continual learning is completed to make the model fit for multiple fusion tasks.
	In U2Fusion~\cite{9151265}, VGG-16~\cite{simonyan2014very} is exploited as a feature extractor to capture multi-level features, and then the importance of related information is determined by adaptive information preservation degrees. Finally, the DenseNet module~\cite{HuangLMW17} is trained to retain the adaptive similarity between the fusion/source images.
	For the third group, Li et al.~\cite{li2018infrared} carry out the two-scale decomposition through the optimization method in section~\ref{ts_de}, then the base part is merged through a weighted-averaging strategy, and the fused detail content is obtained by a pretrained VGG-19~\cite{simonyan2014very} network cooperating with the softmax operator. The final fusion image is reconstructed from base and detail feature maps.
	In literature~\cite{lahoud2019fast}, Lahoud and S{\"u}sstrunk use the blur filters to decompose images. Whereafter, the base part is fused through a saliency weighting strategy, and the feature extraction of the detail part is completed through ResNet50~\cite{HeZRS16}.	
	For the AE-based group, Li and Wu~\cite{li2018densefuse} separate the fusion process into two stages. In the training stage, the useful source image features are extracted by training an AE structure based on Densenet~\cite{HuangLMW17}. In the test stage, after merging the feature maps output by the encoder, the fused images are acquired by the well-trained decoder.
	
	\section{Method}\label{sec:3}
	In this section, we will introduce our DIDFuse network and its structure. Details of training and testing phases are also illustrated.
	
	\subsection{Motivation}
	As described in section \ref{ts_de}, two-scale decomposition decomposes the input image into a base image containing low-frequency information and a detail image embodying high-frequency information.
	Currently, most algorithms incorporate certain prior knowledge, and employ filters or optimization-based methods to decompose images. Hence, they are manually designed decomposition algorithms. We highlight that image decomposition algorithms are intrinsically feature extractors. Formally, they transform source images from the spatial domain into the feature domain. It is well known that the DNN is a promising data-driven feature extractor and has great superiority over traditional manually-designed methods. Unfortunately, it lacks a DL-based image decomposition algorithm for IVIF task.
	
	Consequently, we present a novel deep image decomposition network in which an encoder is exploited to perform two-scale decomposition and extract different types of information, and a decoder is used to recover original images.
	
	\begin{figure*}[t]
		\centering
		\includegraphics[width=0.8\linewidth]{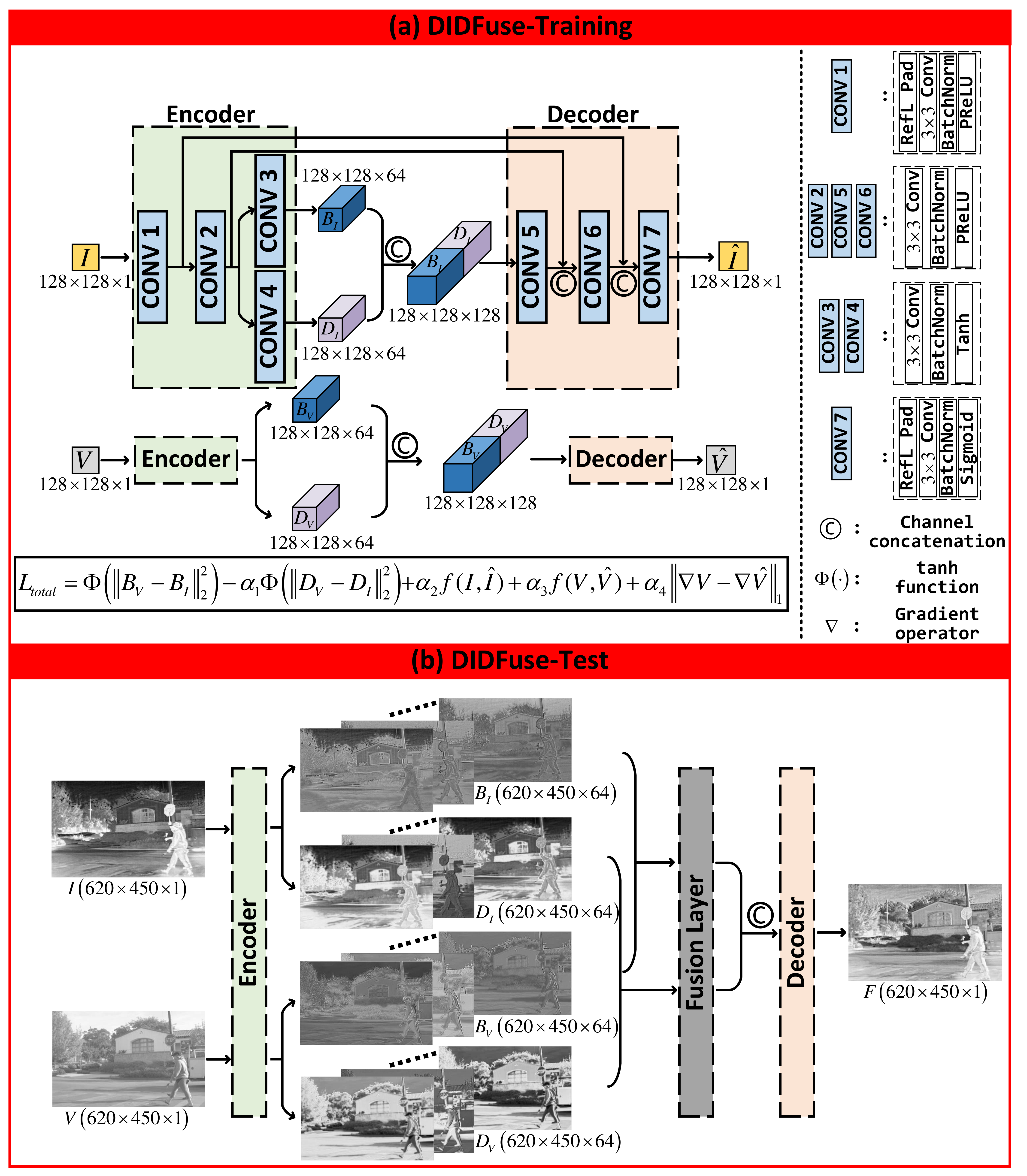}
		\caption{Neural network framework of DIDFuse.}
		\label{DIDF}
	\end{figure*}
	\begin{figure*}[t]
		\centering
		\includegraphics[width=\linewidth]{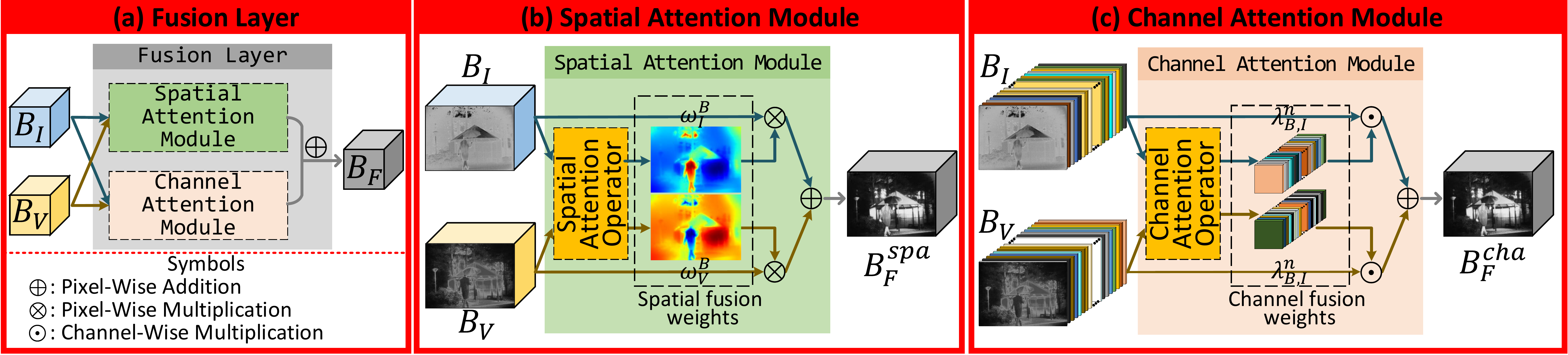}
		\caption{The illustration of fusion layer. We only show the merging process of the base feature maps, and that of the detail feature maps is very similar to it by substituting $\left\{B_I,B_V,B_F\right\}$ for $\left\{D_I,D_V,D_F\right\}$.}
		\label{fig:fusionLayer}
	\end{figure*}	
	\subsection{Network Architecture}
	Our neural network consists of an encoder and a decoder. As illustrated in Figure \ref{DIDF}, the encoder is fed with an infrared or a visible image and generates base and detail feature maps. Then, the network concatenates two kinds of feature maps along channels. At last, concatenated feature maps pass through decoder to recover the original image. To prevent the detail information of the feature maps from being lost after multiple convolutions and to speed up the convergence, we add the feature maps from the first and second convolutions to the inputs of the last and penultimate convolutions, and the adding strategy is concatenating the corresponding feature maps along channels. As a consequence, the pixel intensity and gradient information of the source images can be better retained in the reconstructed image.
		
	\begin{table}[tbp]
		\centering
		\caption{Network configuration. Size denotes the size of convolutional kernel. InC and OutC are the numbers of input and output channels, respectively.}
		\begin{tabular}{cccccc}
			\toprule
			Layers & Size & InC & OutC &  Padding   & Activation \\ \midrule
			conv1  &  3   &  1  &  64  & Reflection &   PReLU    \\
			conv2  &  3   & 64  &  64  &    Zero    &   PReLU    \\
			conv3  &  3   & 64  &  64  &    Zero    &    Tanh    \\
			conv4  &  3   & 64  &  64  &    Zero    &    Tanh    \\
			conv5  &  3   & 128 &  64  &    Zero    &   PReLU    \\
			conv6  &  3   & 64  &  64  &    Zero    &   PReLU    \\
			conv7  &  3   & 64  &  1   & Reflection &  Sigmoid   \\ \bottomrule
		\end{tabular}%

		\label{tab:net_configuration}%
	\end{table}%
	
	Table \ref{tab:net_configuration} lists the network configuration. Encoder and decoder contain four and three convolutional layers, respectively. Each layer consists of a padding, a $3\times3$ convolution, a batch normalization and an activation function. The first and the last layers utilize reflection padding to prevent artifacts at the edges of the fused image. Activation functions of \texttt{conv3} and \texttt{conv4} are set to the hyperbolic tangent function (tanh) since they output base and detail feature maps. As for \texttt{conv7}, it is activated by sigmoid function since it reconstructs original images. Other layers are followed by parametric rectified linear units (PReLU).
	
	\subsection{Loss Function}
	In the training phase, we aim to obtain an encoder that performs two-scale decomposition on the source images, and at the same time, acquire a decoder that can fuse the images and preserve the information of source images well. The training process is shown in Figure \ref{DIDF}(a).
	
	\subsubsection{Feature decomposition}Base feature maps are used to extract the common features of source images, while detail feature maps are used to capture the distinct characteristics from infrared and visible images. Therefore, we should make the gap of base feature maps small. In contrast, the gap of detail feature maps should be great. To this end, the loss function of feature decomposition is defined as follows,
	\begin{equation}\label{L1}
	{L_1} = \Phi \left( {{{\left\| {{B_V} - {B_I}} \right\|}_2^2}} \right) - {\alpha _1}\Phi \left( {{{\left\| {{D_V} - {D_I}} \right\|}_2^2}} \right),
	\end{equation}
	where $B_V$, $D_V$ are the base and detail feature maps of the visible image $V$, and $B_I$, $D_I$ are those of the infrared image $I$. $\Phi \left(  \cdot  \right)$ is tanh function that is used to bound gap into interval $(-1,1)$.
	
	\subsubsection{Image Reconstruction}
	As for image reconstruction, to successfully retain the pixel intensity and detailed texture information of input images, the reconstruction loss function is given by
	\begin{equation}\label{L2}
	{L_2} = {\alpha_2}f(I,\hat I) + {\alpha _3}f(V,\hat V) + {\alpha _4}{\left\| {\nabla V - \nabla \hat V} \right\|_1},
	\end{equation}
	where $I$ and $\hat I$, $V$ and $\hat V$ represent the input and reconstructed images of infrared and visible images, respectively. $\nabla$ denotes the gradient operator, and
	\begin{equation}\label{fxxh}
	f(X,\hat X) = \left\| {X - \hat X} \right\|_2^2 + \lambda {L_{SSIM}}(X,\hat X),
	\end{equation}
	where $X$ and $\hat X$ represent the above input image and the reconstructed image, and $\lambda$ is the hyperparameter. SSIM is the structural similarity index\cite{wang2004image}, which is a measure of the similarity between two pictures. Then $L_{SSIM}$ can be described as 
	\begin{equation}\label{}
	{L_{SSIM}}(X,\hat X) = \frac{{1 - SSIM(X,\hat X)}}{2}.
	\end{equation}

	Remark that $\ell_2$-norm measures the pixel intensity agreement between original and reconstructed images, and that $L_{SSIM}$ computes image dissimilarity in terms of brightness, contrast and structure. Specially, since visible images are with enriched textures, the reconstruction of visible images is regularized by gradient sparsity penalty to guarantee texture agreement.
	
	Combining Eqs.~(\ref{L1}) and (\ref{L2}), the total loss $L_{total}$ can be expressed as
	\begin{equation}\label{totalloss}
	\begin{split}
	L_{total} =& L_1+L_2 \\
	=& \Phi \left( {\left\| {{B_V} - {B_I}} \right\|_2^2} \right) - {\alpha _1}\Phi \left( {\left\| {{D_V} - {D_I}} \right\|_2^2} \right) \\
	+& {\alpha _2}f(I,\hat I) + {\alpha _3}f(V,\hat V) + {\alpha _4}{\left\| {\nabla V - \nabla \hat V} \right\|_1},
	\end{split}
	\end{equation}
	where ${\alpha _1},{\alpha _2},{\alpha _3},{\alpha _4}$ are the tuning parameters.

	\subsection{Fusion Strategy}\label{ADS}
	In the above subsections, we have proposed network structure and loss function. After training, we will acquire a decomposer (or say, encoder) and a decoder. In the test phase, we aim to fuse infrared and visible images. The workflow is shown in Figure~\ref{DIDF}(b). Different from training, a fusion layer is inserted in the test phase. It merges base and detail feature maps separately. In formula, there is
	\begin{equation}
	B_F = {\rm Fusion}(B_I,B_V), D_F = {\rm Fusion}(D_I,D_V),
	\end{equation}
	where $B_F$ and $D_F$ denote the fused base and detail feature maps, respectively. In this paper, the fusion layer can be devided into two parts: spatial attention modual~(SAM) and channel attention module~(CAM), where SAM focuses on the activity degree for each pixel and CAM extracts the importance degree of each channel in feature maps for fusion tasks. The illustration of fusion layer is displayed in Figure~\ref{fig:fusionLayer}. Due to space constraints, we only show the merging process of the base feature maps, and that of the detail feature maps is very similar to it by substituting $\left\{B_I,B_V,B_F\right\}$ for $\left\{D_I,D_V,D_F\right\}$.\\
	\textbf{Spatial attention module:} 
	three fusion strategies are considered in this module as follows:\\
	(1) \textit{$\ell_1$-attention average strategy:}
	Referring to \cite{li2018densefuse}, we use the $\ell_1$-norm as a measure of activity, combining with the softmax operator.
	In detail, we can obtain the activity level map of the fused base and detail feature maps by ${\left\| {{B_i}(x,y)} \right\|}_1$ and ${\left\| {{D_i}(x,y)} \right\|}_1 (i = 1,2)$, where $B_1$, $B_2$, $D_1$ and $D_2$ represent $B_I$, $B_V$, $D_I$ and $D_V$, and $(x,y)$ represents the corresponding coordinates of the feature maps.
	Then the merging weights can be calculated by:
	\begin{equation}\label{}
	\begin{split}
	\eta _i^B(x,y) = &  \frac{{\psi \left( {{{\left\| {{B_i}(x,y)} \right\|}_1}} \right)}}{{\sum\nolimits_{i = 1}^2 {\psi \left( {{{\left\| {{B_i}(x,y)} \right\|}_1}} \right)} }} , \\
	\eta _i^D(x,y) = &  \frac{{\psi \left( {{{\left\| {{D_i}(x,y)} \right\|}_1}} \right)}}{{\sum\nolimits_{i = 1}^2 {\psi \left( {{{\left\| {{D_i}(x,y)} \right\|}_1}} \right)} }},
	\end{split}
	\end{equation}
	where $\psi (\cdot)$ is a $3\times3$ box blur (also known as a mean filter operator). Consequently, we have
	\begin{equation}\label{}
	\begin{split}
	B_F^{spa} & = (\eta _1^B\otimes{B_I}) \oplus (\eta _2^B\otimes{B_V}), \\
	D_F^{spa} & = (\eta _1^D\otimes{D_I})\oplus (\eta _2^D\otimes{D_V}).
	\end{split}
	\end{equation}
	where $B_F^{spa}$ and $D_F^{spa}$ denote the merging feature map of base and detail feature maps after SAM, respectively. $\otimes$ means pixel-wise multiplication and $\oplus$ denotes pixel-wise addition.\\	
	(2) \textit{Saliency-attention average strategy:}
	Inspired by \cite{DBLP:journals/corr/abs-1905-03590,DBLP:conf/mm/ZhaiS06}, the saliency degree is used to determine the merging weights for highlighting the objects and preserving more information from source images. We take the calculation of base feature map weights as an example.
	Firstly, the saliency degree $\mathcal{S}_I^B$ of $B_I$ at pixel $\left(x,y\right)$ can be obtained by:
	\begin{equation}\label{equ:saliency1}
	\mathcal{S}_I^B\left(x,y\right)=\sum_{k=0}^{255} H^B(i)|B_I\left(x,y\right)-i|,
	\end{equation}
	where $B_I\left(x,y\right)$ is the image intensity of $B_I$ at pixel~$(x,y)$ and $H^B_I(k)|k=1,\ldots,255$ is the histogram of $B_I$. $\mathcal{S}_V^B$ and $H^B_V$ for $B_V$ can be acquired similarly.
	Then we obtain the saliency-based weight maps at pixel $\left(x,y\right)$ by
	\begin{equation}\label{}
	{\tilde\omega_I^B\left(x,y\right)} \!=\! \frac{{\mathcal{S}_I^B\left(x,y\right)}}{{\mathcal{S}_I^B\left(x,y\right)\! +\! \mathcal{S}_V^B\left(x,y\right)}}, \tilde\omega _V^B\!\left(x,y\right)\!=\!1\!-\!\tilde\omega_I^B\!\left(x,y\right).
	\end{equation}
	Afterwards, guided filtering~$\chi(\cdot,\cdot)$ is employed to solve the spatial inconsistency issue, such as the clear boundaries around objects and artifacts:
	\begin{equation}\label{equ:saliency2}
	{\omega_I^B} = \frac{{\chi({{\tilde \omega_I}^B},{B_I})}}{{\chi({{\tilde \omega_I}^B},{B_I}) + \chi({{\tilde \omega_V}^B},{B_V})}},{\omega_V^B}=1-{\omega_I^B},
	\end{equation}
	and the merging feature map $B_F^{spa}$ is acquired by
	\begin{equation}\label{equ:sal_bf}
		B_F^{spa} = (\omega_I^B\otimes{B_I}) \oplus (\omega_V^B\otimes{B_V}),
	\end{equation}
	where the symbol $\oplus$ means element-wise addition. More calculation details can be referred in \cite{DBLP:conf/mm/ZhaiS06}.\\
	\textit{(3) Weighted-average strategy:}
	As a control group, we give a manually designed merging weight, i.e., 
	\begin{equation}\label{}
	B_F^{spa} = {\gamma _1}{B_I} \oplus {\gamma _2}{B_V}, D_F^{spa} = {\gamma _3}{D_I} \oplus {\gamma _4}{D_V},
	\end{equation}
	where ${\gamma _1}+{\gamma _2={\gamma_3}+{\gamma_4}=1}$ and the default settings for ${\gamma _i}(i=1,\cdot\cdot\cdot,4)$ are all equal to 0.5.\\
	\textbf{Channel attention module:} For traditional feature fusion algorithms, most of works~\cite{li2018densefuse,li2018infrared,liu2016image} pay more attention to the merging weights of the spatial domain, and ignore that of the channel dimension. Inspired by \cite{DBLP:conf/eccv/WooPLK18,DBLP:journals/tim/LiWD20}, we establish this module to calculate the channel-wise merging weight based on the channel attention mechanism, which highlights more important conspicuous features from channel perspective and retains more features for specific channels. Similarly with SAM, we take base feature maps $\{B_I,B_V\}$ as examples to show the calculation details. 
	
	Firstly, we obtain the activity degree with respect to each channel of $B_I$ or $B_V$ by the global pooling operator~$\Omega(\cdot)$, i.e.,
	\begin{equation}\label{}
	\tilde\lambda_{B,I}^n = \Omega(B_I^n), \tilde\lambda_{B,V}^n = \Omega(B_V^n),
	\end{equation}
	where $\left\{B_I^n,B_V^n\right\}$ represent the $n$th channel of $\left\{B_I,B_V\right\}$ $(n=1,2,\cdots,N)$, and $\{\tilde\lambda_{B,I}^n,\tilde\lambda_{B,V}^n\}$ denote the initial activity degree of $\left\{B_I^n,B_V^n\right\}$, respectively. Then $\tilde\lambda_{B,I}^n$ and $\tilde\lambda_{B,V}^n$ are normalized to get the final merging weight vector by
	\begin{equation}\label{}
	\lambda_{B,I}^n=\frac{\tilde\lambda_{B,I}^n}{\tilde\lambda_{B,I}^n+\tilde\lambda_{B,V}^n},\lambda_{B,V}^n=1-\lambda_{B,I}^n.
	\end{equation}
	Ultimately, the merged feature maps $B_F^{cha}$ output by CAM can be acquired by
	\begin{equation}\label{}
	B_F^{cha}=\left( \lambda_{B,I}\odot {B_I}\right) \oplus \left( \lambda_{B,V}\odot {B_V}\right) 
	\end{equation}
	where $\odot$ denotes channel-wise multiplication. 
	
	At last, the merged feature maps output by the total fusion layer are acquired by
	\begin{equation}\label{}
	B_F\!=\!\left(B_F^{spa}\!+\!B_F^{cha}\right)\!\times\!0.5,\ D_F\!=\!\left(D_F^{spa}\!+\!D_F^{cha}\right)\!\times\!0.5.
	\end{equation}
	The validation experiments in section~\ref{Va} determine the fusion layer settings, including the selection of merging strategies in SAM and whether to employ the CAM.	
	\subsection{Comparison with related methods}
	Our method combines the characteristics of the second and the third group of DL-based methods in section~\ref{sec:2_3}. Notably, this is the first time that the DL technology is used in the two-scale decomposition. Meanwhile, we transform the fusion task from the image domain into the feature domain through the AE structure.
	Additionally, our data-driven model should not be regarded as a simple extension of the traditional model, which usually perform image decomposition by simple filters or extract feature by pre-trained networks. Different from them, we perform image decomposition and feature extraction by training a designed network and learning the model parameters by a novel loss function.
	
	\section{Experiment} \label{sec:4}
	The aim of this section is to study the performance of our proposed model and compare it with other SOTA models, including FusionGAN~\cite{ma2019fusiongan}, Densefuse~\cite{li2018densefuse}, ImageFuse~\cite{li2018infrared}, DeepFuse~\cite{prabhakar2017deepfuse}, TSIFVS~\cite{bavirisetti2016two}, TVADMM~\cite{guo2017infrared}, U2Fusion~\cite{9151265} and NestFuse~\cite{DBLP:journals/tim/LiWD20}. All experiments were conducted with Pytorch on a computer with Intel Core i9-10900K CPU@3.70GHz and NVIDIA GeForce RTX2080Ti GPU.
	
	\subsection{Experiment preparation}	
	\subsubsection{Datasets and preprocessing}
	Our experiments are conducted on three datasets, including TNO~\cite{TNO}, NIR~\cite{brown2011multi} and FLIR~\cite{xu2020aaai}.
	In our experiment, we divide them into training, validation, and test sets. Table~\ref{dataset} shows the numbers of image pairs, illumination and scene information of the datasets.
	We randomly selected 180 pairs of images in the FLIR dataset and two scenery in NIR dataset as training samples. Before training, all images are transformed into grayscale. At the same time, we center-crop them with $128\times128$ pixels. It is worth noting that the center-crop operation only appears in the training phase rather than the test phase.
	\begin{table}[tbp]
		\centering
		\caption{Dataset used in this paper.}
		\label{dataset}
		\begin{tabular}{ccc}
			\toprule
			                          &  Dataset(pairs)  &     Illumination     \\ \midrule
			 \multirow{3}*{Training}  & FLIR-Train(180)  & Daylight\&Nightlight \\
			                          & NIR-Mountain(55) &       Daylight       \\
			                          &  NIR-Water(51)   &       Daylight       \\ \midrule
			\multirow{2}*{Validation} &  NIR-Urban(58)   &       Daylight       \\
			                          &  NIR-Street(50)  &       Daylight       \\ \midrule
			   \multirow{5}*{Test}    &     TNO	(40)     &      Nightlight      \\
			                          &  FLIR-Test(40)   & Daylight\&Nightlight \\
			                          & NIR-Country(52)  &       Daylight       \\
			                          &  NIR-Forest(53)  &       Daylight       \\
			                          &  NIR-Field(51)   &       Daylight       \\ \bottomrule
		\end{tabular}
	\end{table}
	
	\begin{figure}[tb]
		\centering
		\includegraphics[width=1\linewidth]{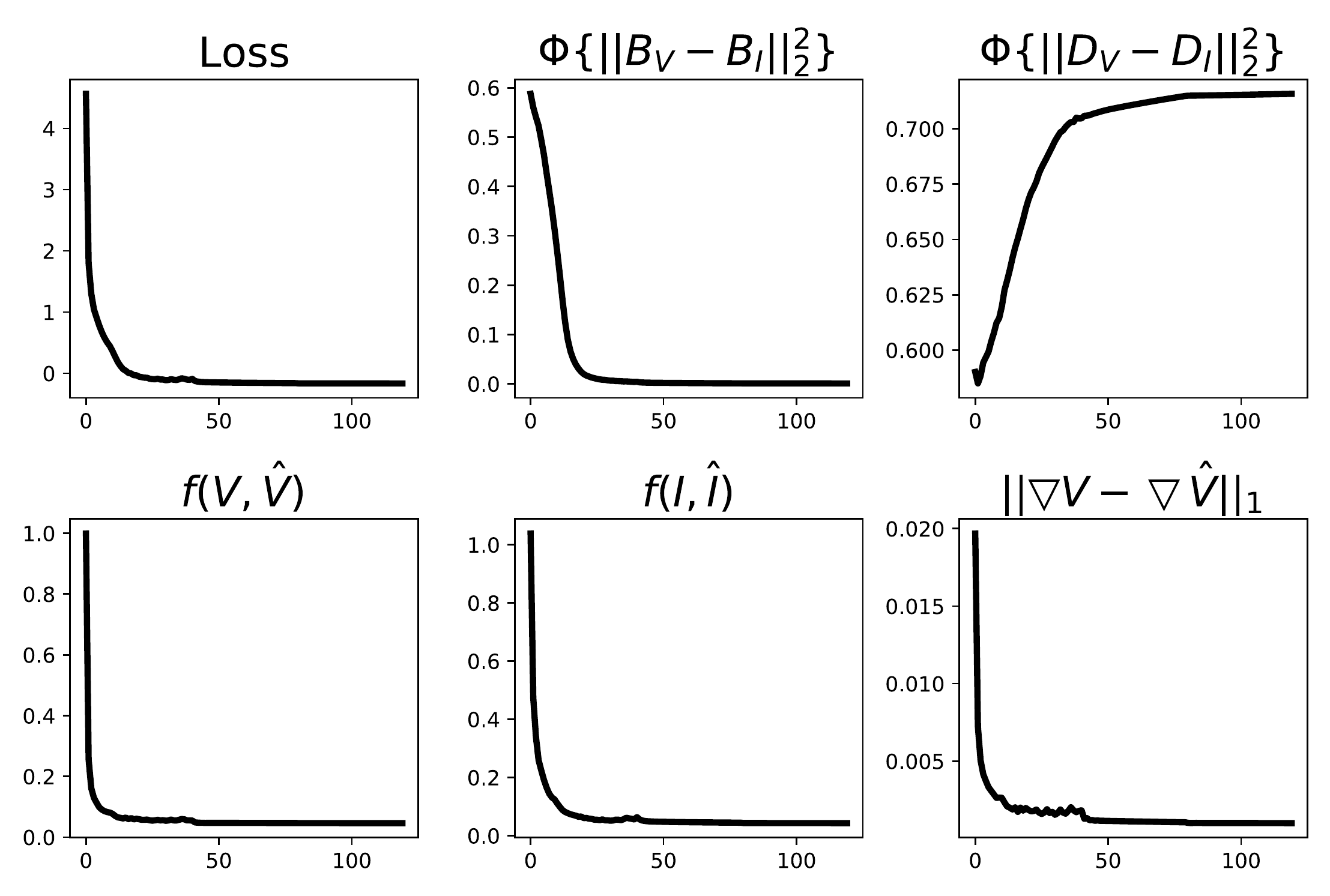}
		\caption{Loss curves over 120 epochs.}
		\label{Loss}
	\end{figure}
	\begin{table*}[tbp]
		\centering
		\caption{Results of two validation sets for choosing the merging strategy and the largest value is shown in bold. From left to right: weighted-average strategy w/o CAM, $\ell_1$-attention average strategy w/o CAM, saliency-attention average strategy w/o CAM, weighted-average strategy with CAM, $\ell_1$-attention average strategy with CAM and saliency-attention average strategy with CAM.}
		\label{table_VAL}
		\begin{tabular}{crrrrrr}
			\toprule
			                                                                                     \multicolumn{7}{c}{\textbf{Dataset: NIR Dataset. Scene: Street}}                                                                                      \\
			Metrics & \makebox[1.7cm][c]{Ave w/o CAM} & \makebox[1.7cm][c]{$\ell_1$-Att w/o CAM} & \makebox[1.7cm][c]{Sal-Att w/o CAM} & \makebox[1.7cm][c]{Ave \& CAM} & \makebox[1.7cm][c]{$\ell_1$-Att \& CAM} & \makebox[1.7cm][c]{Sal-Att \& CAM} \\ \midrule
			  EN    &              7.144 	$\pm$	0.023 &                       7.147 	$\pm$	0.036 &                  7.183 	$\pm$	0.033 &             7.195 	$\pm$	0.069 &                      7.196 	$\pm$	0.073 &        \textbf{7.214 	$\pm$	0.068} \\
			  SD    &             48.360 	$\pm$	0.956 &                      48.531 	$\pm$	1.223 &                 49.451 	$\pm$	1.253 &            56.613 	$\pm$	1.289 &                     56.699 	$\pm$	1.323 &       \textbf{57.288 	$\pm$	1.436} \\
			  SF    &             21.160 	$\pm$	0.470 &                      21.186 	$\pm$	0.533 &                 21.342 	$\pm$	0.484 &            24.943 	$\pm$	0.736 &                     24.958 	$\pm$	0.762 &       \textbf{25.059 	$\pm$	0.732} \\
			  VIF   &              0.886 	$\pm$	0.015 &                       0.886 	$\pm$	0.022 &                  0.895 	$\pm$	0.018 &             1.042 	$\pm$	0.025 &                      1.042 	$\pm$	0.027 &        \textbf{1.050 	$\pm$	0.025} \\
			  AG    &              6.331 	$\pm$	0.147 &                       6.340 	$\pm$	0.172 &                  6.400 	$\pm$	0.144 &             7.397 	$\pm$	0.271 &                      7.401 	$\pm$	0.283 &        \textbf{7.441 	$\pm$	0.263} \\
			  SCD   &              1.539 	$\pm$	0.035 &                       1.536 	$\pm$	0.040 &                  1.540 	$\pm$	0.042 &    \textbf{1.702 	$\pm$	0.042} &                      1.699 	$\pm$	0.041 &                 1.692 	$\pm$	0.044 \\ \midrule
			                                                                                     \multicolumn{7}{c}{\textbf{Dataset: NIR Dataset. Scene: Urban}}                                                                                       \\
			Metrics & \makebox[1.7cm][c]{Ave w/o CAM} & \makebox[1.7cm][c]{$\ell_1$-Att w/o CAM} & \makebox[1.7cm][c]{Sal-Att w/o CAM} & \makebox[1.7cm][c]{Ave \& CAM} & \makebox[1.7cm][c]{$\ell_1$-Att \& CAM} & \makebox[1.7cm][c]{Sal-Att \& CAM} \\ \midrule
			  EN    &              7.284 	$\pm$	0.034 &                       7.284 	$\pm$	0.037 &         \textbf{7.302 	$\pm$	0.036} &             7.191 	$\pm$	0.097 &                      7.191 	$\pm$	0.098 &                 7.202 	$\pm$	0.097 \\
			  SD    &             54.134 	$\pm$	0.767 &                      54.209 	$\pm$	0.891 &                 54.789 	$\pm$	0.917 &            62.345 	$\pm$	1.213 &                     62.385 	$\pm$	1.239 &       \textbf{62.778 	$\pm$	1.276} \\
			  SF    &             25.510 	$\pm$	0.482 &                      25.524 	$\pm$	0.500 &                 25.596 	$\pm$	0.485 &            29.825 	$\pm$	1.021 &                     29.833 	$\pm$	1.029 &       \textbf{29.884 	$\pm$	1.015} \\
			  VIF   &              1.019 	$\pm$	0.014 &                       1.019 	$\pm$	0.015 &                  1.026 	$\pm$	0.015 &             1.140 	$\pm$	0.028 &                      1.140 	$\pm$	0.029 &        \textbf{1.145 	$\pm$	0.028} \\
			  AG    &              7.353 	$\pm$	0.144 &                       7.357 	$\pm$	0.153 &                  7.390 	$\pm$	0.143 &             8.314 	$\pm$	0.320 &                      8.315 	$\pm$	0.325 &        \textbf{8.340 	$\pm$	0.316} \\
			  SCD   &              1.577 	$\pm$	0.028 &                       1.574 	$\pm$	0.029 &                  1.582 	$\pm$	0.032 &             1.675 	$\pm$	0.046 &                      1.673 	$\pm$	0.045 &        \textbf{1.674 	$\pm$	0.043} \\ \bottomrule
		\end{tabular}
	\end{table*}
	\subsubsection{Evaluation metrics}
	As an unsupervised task, there are no ground truth images for reference in the above fusion datasets. So we employ six metrics to evaluate the quality of a fused image, that is,
	entropy (EN), standard deviation (SD), spatial frequency (SF), visual information fidelity (VIF), average gradient (AG) and sum of the correlations of differences (SCD). The introduction and calculation details are as follows:
	\begin{itemize}
		\item Entropy (EN): It reflects the amount of overall information contained in an image, and it is defined by
		\begin{equation}
		{\rm EN} = -\sum_{l=0}^{255} p_l\log_2 p_l,
		\end{equation}
		where $p_l$ is the normalized frequency of the corresponding gray level in an image. The larger the EN is, the more information is contained in an image.
		\item Standard deviation (SD): It is defined by
		\begin{equation}
		{\rm SD} = \sqrt{\frac{1}{hw}\sum_{i,j} (I_{ij}-\mu)^2},
		\end{equation}
		where $I_{ij}$ is the $(i,j)$ pixel value in the fused image $I$, and $\mu$ denotes mean pixel value. To some extent, larger SD indicates that an image is with high contrast, providing better visual effect.
		\item Spatial frequency (SF): SF is a gradient-based image quality metric, which measures the horizontal gradients~(HG) and vertical gradients~(VG) of the input image to reveal the details and texture of the image with the definition as follows
		\begin{equation}\label{}
		SF=\sqrt{HG^2+VG^2},
		\end{equation}
		where
		\begin{equation}\label{}
		\begin{split}
		HG = & \sqrt{\textstyle \sum_{i} \sum_{j}(I(i, j)-I(i, j-1))^{2}}, \\
		VG = & \sqrt{\textstyle \sum_{i} \sum_{j}\left(I(i, j)-I(i-1, j)\right)^{2}}.
		\end{split}
		\end{equation}
		The larger SF value means the richer edges and texture details containing in the fused image.
		\item Visual information fidelity (VIF): VIF measures the fusion performance of images by calculating the fidelity of visual information.
		Firstly, the source image and the fused image are divided into blocks by wavelet decomposition.
		Then, mutual information is calculated for each block and band to evaluate the degree of information distortion.
		Finally, the visual information of all blocks and all bands are integrated to measure the image quality.
		The larger the VIF value is, the better the fusion image can meet the human visual system.
		\item Average gradient (AG): AG indicates the details and texture information of the fused image by calculating the gradient information of the fused image in the horizontal and vertical directions, which is defined as follows:
		\begin{equation}\label{}
		AG=\frac{1}{MN} \sum_{i} \sum_{j} \sqrt{\frac{\nabla I_{h}^{2}(i, j)+\nabla I_{v}^{2}(i, j)}{2}}
		\end{equation}
		where
		\begin{equation}\label{}
		\begin{split}
		\nabla I_{h}(i, j) = & I(i, j)-I(i, j-1), \\
		\nabla I_{v}(i, j) = & I(i, j)-I(i-1, j).
		\end{split}
		\end{equation}
		The larger the AG value is, the more texture and detail information containing in the fused image.
		\item Sum of the correlations of differences (SCD):
		This metric measures the impact of source images on the fusion image, based on the correlation operation. It is defined by:
		\begin{equation}\label{}
		SCD=\phi(F-S_V,S_I)+\phi(F-S_I,S_V),
		\end{equation}
		where $F$ is the fusion image, $S_I$ and $S_V$ represent the infrared and visible images, respectively. $\phi(\cdot,\cdot)$ denotes the correlation operator, which is calculated by:
		\begin{equation}
		\phi \left(I_{1}, I_{2}\right)=\frac{\sum\left(I_{1}-\bar{I}_{1}\right)\left(I_{2}-\bar{I}_{2}\right)}{\sqrt{\left(\sum\left(I_{1}-\bar{I}_{1}\right)^{2}\right)\left(\sum\left(I_{2}-\bar{I}_{2}\right)^{2}\right)}}.
		\end{equation}	
		The larger SCD is, the more information is transferred from source images to the fused image.
	\end{itemize}
	In short, the above metrics evaluated fusion images from different aspects. Therefore, in the experiment, we need to comprehensively consider the values of them to compare the performance of different fusion methods. And more details for these metrics can be seen in \cite{ma2019infrared}.

	\subsubsection{Hyperparameters setting}
	As we know, best hyperparameters can be found with grid searching in validation datasets, but it is time-consuming.
	So in our model, the tuning parameters in loss function are empirically set as follows:  ${\alpha _1}=0.05$, ${\alpha _2}= 2$, ${\alpha _3}=2$, ${\alpha _4}=10$ and $\lambda=5$. For $\alpha_2$ to $\alpha_4$ and $\lambda$, we keep the values of the loss items with the same magnitude, and for $\alpha_1$, experiments show that other loss items decline slowly and the model is not easy to be trained if it is set to a too large value, so a relatively small weight is preferred.
	
	Moreover, in training phase, the network is optimized by Adam over 120 epochs with a batch size of 24.
	As for the learning rate, we set it to $10^{-3}$ and decrease it by 10 times every 40 epochs.
	Figure \ref{Loss} displays loss curves versus epoch index. It is shown that all loss curves are very flat after 120 epochs. In other words, the network is able to converge with this configuration.
	
	\begin{figure*}[h]
	\centering
	\includegraphics[width=0.8\linewidth]{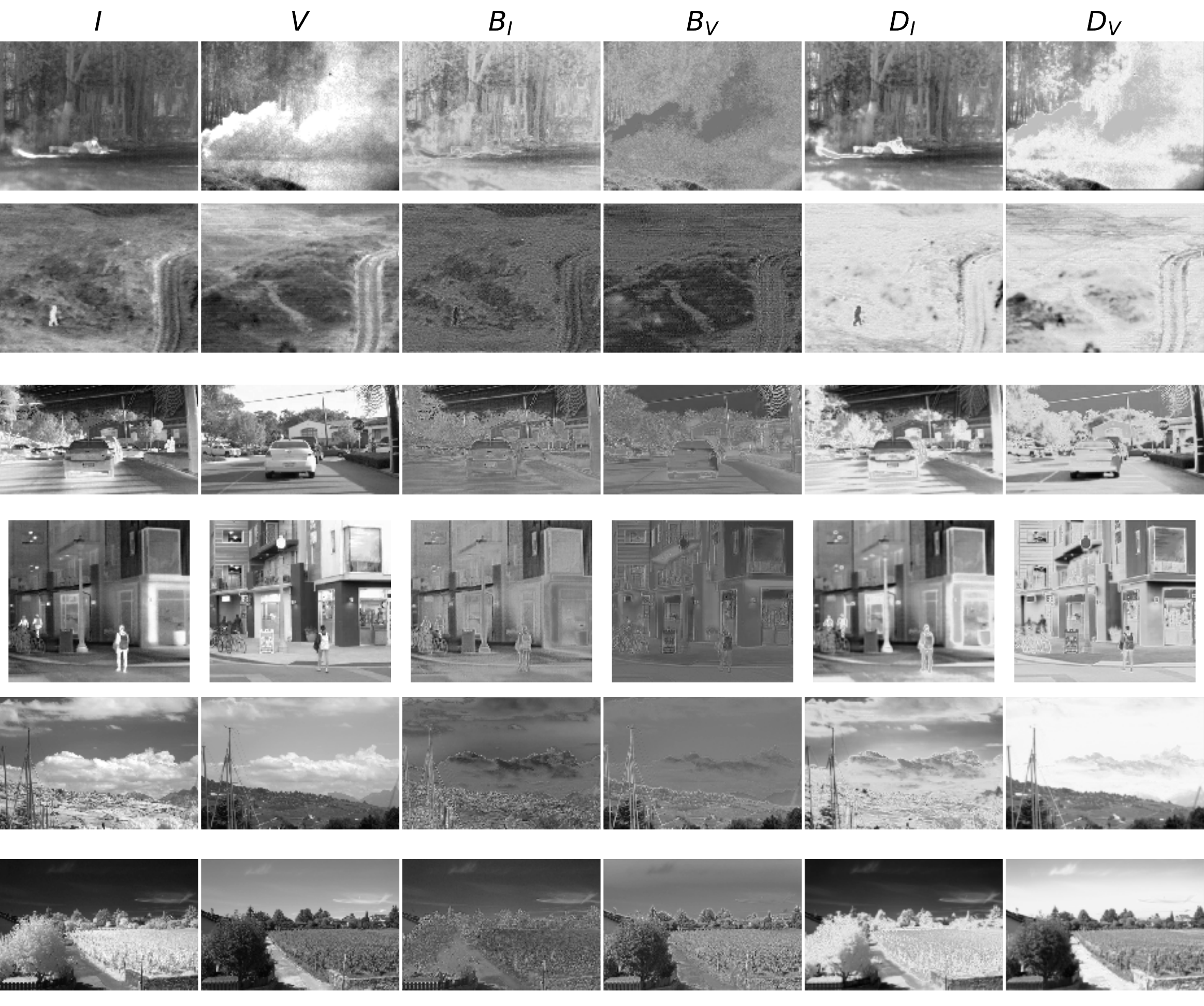}
	\caption{Illustration of deep image decomposition. From left to right: infrared image, visible image, base and detail feature maps of infrared image and visible image. }
	\label{base_detail}
\end{figure*}	
	\subsection{Experiments on Fusion Strategy}\label{Va}
	As described in section \ref{ADS}, merging strategy in fusion layer plays an important role in our model. We investigate the performance of six strategies on validation set and the numerical results of six metrics are reported in Table \ref{table_VAL}. 
	Compared with the three strategies containing only the SAM, the other three strategies combining CAM and SAM have better performance, demonstrating the effectiveness of the channel attention mechanism in CAM. At the same time, regardless of whether the CAM module is employed, the saliency-attention average strategy has higher metrics than $\ell_1$-attention and weighted-average strategies.
	Obviously, it is shown that \textit{saliency-attention average strategy} in SAM cooperating with {CAM} achieves higher values, especially in terms of SD, SF, VIF and AG. Hence, the following experiments adopt this merging strategy in the fusion layer.
	\subsection{Experiments on Image Decomposition}
	One of our contributions is the deep image decomposition. It is interesting to study whether decomposed feature maps are able to meet our demands. 
	In Figure~\ref{base_detail}, it displays the first channels of feature maps which are generated by \texttt{conv3} and \texttt{conv4}. It is evident that our method can separate the base backgrounds and detail contents of infrared and visible images. For base feature maps, it is found that $B_I$ and $B_V$ are visually similar, and they reflect the background and environment of the same scene. Conversely, the gap between $D_I$ and $D_V$ is large, which illustrates the distinct characteristics contained in different source images. That is, the infrared images contain target highlight and thermal radiation information while gradient and texture information of targets are involved in the visible images. In conclusion, it verifies the rationality of our proposed network structure and image decomposition loss function to some degree.

	\subsection{Ablation experiments}
	\begin{table*}[!]
		\centering
		\caption{Results of the ablation experiment. In each ablation experiment, the larger value of every metric is shown in bold.}
		\label{table_ablation}
		\begin{tabular}{ccccccccccc}
			\toprule
			                                                                                                           \multicolumn{11}{c}{\textbf{Dataset: NIR Dataset. Scene: Street}}                                                                                                             \\
			                &   \multicolumn{2}{c}{\textbf{Ablation Exp. 1}}    &    \multicolumn{2}{c}{\textbf{Ablation Exp. 2}}     &    \multicolumn{2}{c}{\textbf{Ablation Exp. 3}}     &    \multicolumn{2}{c}{\textbf{Ablation Exp. 4}}    &   \multicolumn{2}{c}{\textbf{Ablation Exp. 5}}    \\
		Metrics &      Ours       &  Test  &      Ours       &      Test       &      Ours       &      Test      &      Ours       &      Test       &  Ours  &  Test  \\ \midrule
EN    & \textbf{7.214}  & 7.041  & \textbf{7.214}  &      6.977      & \textbf{7.214}  &     6.744      & \textbf{7.214}  &      7.084      & \textbf{7.214}  & 7.133  \\
SD    & \textbf{57.288} & 53.004 &     57.288      & \textbf{58.513} & \textbf{57.288} &     45.082     & \textbf{57.288} &     56.805      & \textbf{57.288} & 53.594 \\
SF    & \textbf{25.059} & 22.532 &     \textbf{25.059}      & {25.048} & \textbf{25.059} &     23.010     &     25.059      & \textbf{25.257} & \textbf{25.059} & 24.571 \\
VIF   & \textbf{1.050}  & 0.928  & \textbf{1.050}  &      0.993      & \textbf{1.050}  &     0.730      & \textbf{1.050}  &      0.986      & \textbf{1.050}  & 0.966  \\
AG    & \textbf{7.441}  & 6.525  & \textbf{7.441}  &      7.160      & \textbf{7.441}  &     6.281      & \textbf{7.441}  &      7.189      & \textbf{7.441}  & 7.255  \\
SCD   & \textbf{1.692}  & 1.448  & \textbf{1.692}  &      1.663      & \textbf{1.692}  &     1.384      & \textbf{1.692}  &      1.635      & \textbf{1.692}  & 1.506  \\ \midrule
			                                                                                                            \multicolumn{11}{c}{\textbf{Dataset: NIR Dataset. Scene: Urban}}                                                                                                             \\
			                &   \multicolumn{2}{c}{\textbf{Ablation Exp. 1}}    &    \multicolumn{2}{c}{\textbf{Ablation Exp. 2}}     &    \multicolumn{2}{c}{\textbf{Ablation Exp. 3}}     &    \multicolumn{2}{c}{\textbf{Ablation Exp. 4}}    &   \multicolumn{2}{c}{\textbf{Ablation Exp. 5}}    \\
			    Metrics     &             Ours             &        Test        &              Ours              &        Test        &              Ours              &        Test        &             Ours              &        Test        &             Ours             &        Test        \\ \midrule
		  EN    & \textbf{7.202}  & 7.028  & \textbf{7.202}  &      6.901      & \textbf{7.202}  &     6.797      & \textbf{7.202}  &      7.049      & \textbf{7.202}  & 7.166  \\
SD    & \textbf{62.778} & 57.762 & \textbf{62.778} &     62.344      & \textbf{62.778} &     48.994     & \textbf{62.778} &     61.610      & \textbf{62.778} & 59.011 \\
SF    & \textbf{29.884} & 27.339 &     29.884      & \textbf{30.966} & \textbf{29.884} &     27.682     &     29.884      & \textbf{30.579} & \textbf{29.884} & 29.143 \\
VIF   & \textbf{1.145}  & 1.026  & \textbf{1.145}  &      1.065      & \textbf{1.145}  &     0.827      & \textbf{1.145}  &      1.082      & \textbf{1.145}  & 1.081  \\
AG    & \textbf{8.340}  & 7.494  & \textbf{8.340}  &      8.137      & \textbf{8.340}  &     7.258      & \textbf{8.340}  &      8.182      & \textbf{8.340}  & 8.118  \\
SCD   & \textbf{1.674}  & 1.505  & \textbf{1.674}  &      1.652      &     {1.674}     & \textbf{1.679} &      1.674      & \textbf{1.682}  & \textbf{1.674}  & 1.538  \\ \midrule
			    Overall     & \multirow{2}{*}{\textbf{12}} & \multirow{2}{*}{0} & \multirow{2}{*}{\textbf{	10	}} & \multirow{2}{*}{2} & \multirow{2}{*}{\textbf{	11	}} & \multirow{2}{*}{1} & \multirow{2}{*}{\textbf{	9	}} & \multirow{2}{*}{3} & \multirow{2}{*}{\textbf{12}} & \multirow{2}{*}{0} \\
			\# Larger value &                              &     &&&&&&&               &  \\ \bottomrule
		\end{tabular}
	\end{table*}
	We demonstrate the role of each module in our network (or loss function) through five ablation experiments. The quantitative results can be found in Table~\ref{table_ablation}, and in each ablation experiment, the larger value of every metric is shown in bold. All numerical results were obtained by averaging the metric values in repeatedly training the network 25 times for both our model and experimental groups.
	\subsubsection{The role of the base-scale module}
	We remove the base feature map (i.e., remove \texttt{CONV3} in Fig.~1~(a)) and use only the detail module for fusion. Meanwhile, the loss function of network for training becomes:
	\begin{equation}\label{}
	\begin{split}
	L_{total} =&   - {\alpha _1}\Phi \left( {\left\| {{D_V} - {D_I}} \right\|_2^2} \right)+ {\alpha _2}f(I,\hat I)\\
	+&  {\alpha _3}f(V,\hat V) + {\alpha _4}{\left\| {\nabla V - \nabla \hat V} \right\|_1}.
	\end{split}
	\end{equation}
	
	\subsubsection{The role of the detail-scale module}
	The \texttt{CONV4} and detail feature maps are eliminated and only the base feature map is employed for fusion. Similarly, the loss function for training is:
	\begin{equation}
	\begin{split}
	L_{total} =& \Phi \left( {\left\| {{B_V} - {B_I}} \right\|_2^2} \right) + {\alpha _2}f(I,\hat I)\\
	+&  {\alpha _3}f(V,\hat V) + {\alpha _4}{\left\| {\nabla V - \nabla \hat V} \right\|_1}.
	\end{split}
	\end{equation}
	
	\subsubsection{The role of two-scale decomposition}
	
	We do not remove any certain modules in the architecture, but the base and detail feature maps are not imposed on being similar or dissimilar, with the optimization function:
	\begin{equation}\label{}
	L_{total} ={\alpha _2}f(I,\hat I) + {\alpha _3}f(V,\hat V) + {\alpha _4}{\left\| {\nabla V - \nabla \hat V} \right\|_1}.
	\end{equation}
	
	\subsubsection{Comparison with traditional AE structure}
	
	Our network changes from input infrared and visible images in pairs to randomly input a single source image with eliminating \texttt{CONV4}. Consequently, the network structure becomes a classic AE network with the loss function:
	\begin{equation}\label{}
	\begin{split}
	L_{total} = {\alpha _2}f(X,\hat X) + {\alpha _4}{\left\| {\nabla X - \nabla \hat X} \right\|_1},
	\end{split}
	\end{equation}
	where $X$ and $\hat X$ denote input and reconstructed images.
	
	\begin{figure*}[htbp]
		\centering
		\includegraphics[width=0.95\linewidth]{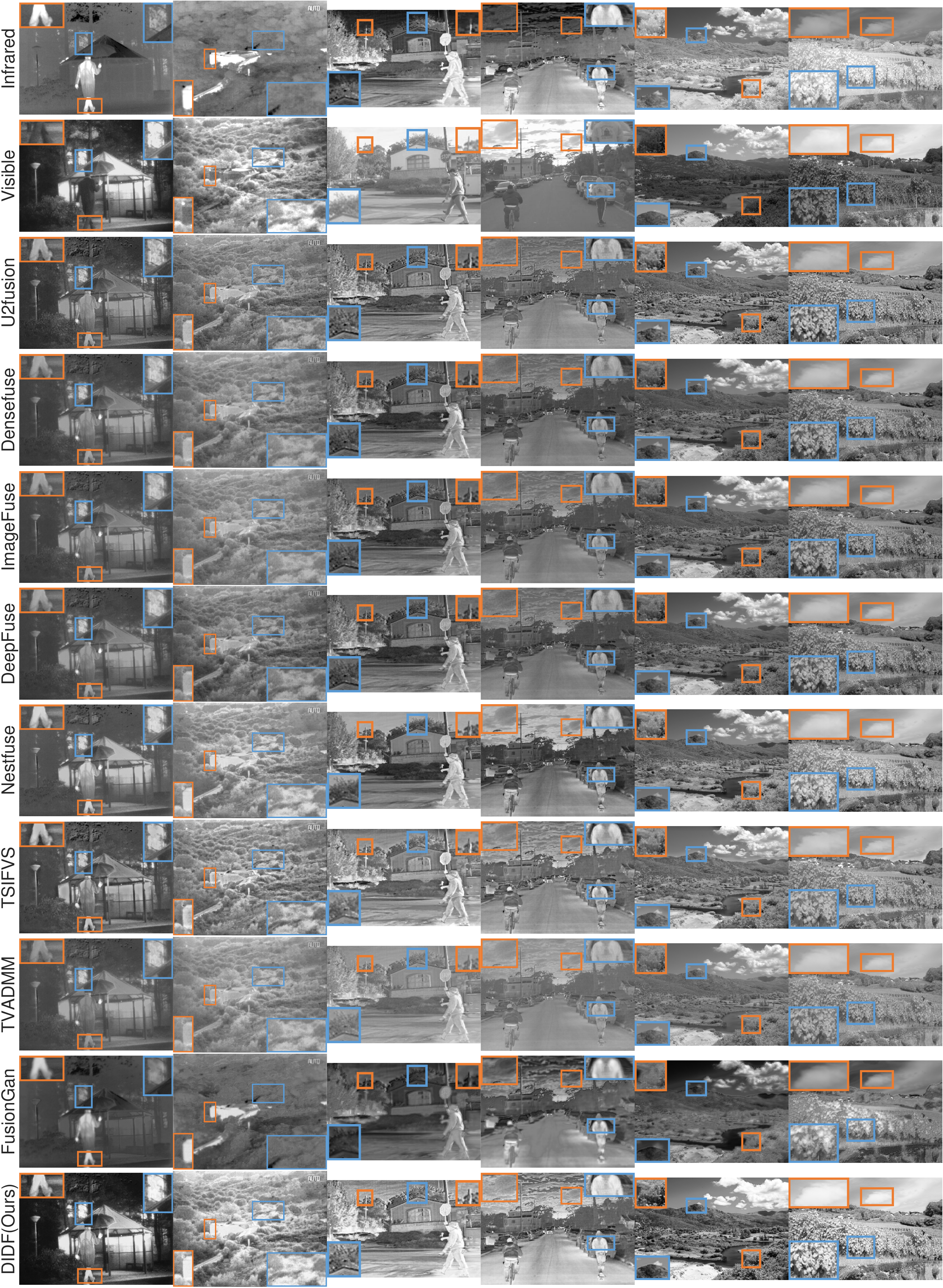}
		\caption{Qualitative results for different methods. Areas marked by orange and blue boxes are amplified for ease of inspection.}
		\label{Figure}
	\end{figure*}	
	\begin{table*}[h]
	\centering
	\caption{Quantitative results of different methods. The largest value is shown in bold, and the second largest value is underlined.}
	\label{table}
	\begin{tabular}{lccccccccc}
		\toprule
		\multicolumn{10}{c}{\textbf{Dataset: TNO Image Fusion Dataset}}                           \\
		Metrics & FusionGAN & DenseFuse & ImageFuse & DeepFuse & TSIFVS & TVADMM & U2fusion & Nestfuse &      DIDFuse       \\ \midrule
		EN      &   6.576   &   6.842   &   6.382   &  6.860   & 6.669  & 6.402  &  6.965   &  \underline{7.025}   &   \textbf{7.107}   \\
		SD      &  29.035   &  31.817   &  22.938   &  32.249  & 28.036 & 23.007 &  33.630  &  \underline{38.501}  &  \textbf{44.251}   \\
		SF      &   8.762   &  11.095   &   9.800   &  11.125  & 12.598 & 9.034  &  13.476  &  \textbf{15.615}  & \underline{14.011} \\
		VIF     &   0.258   &   0.572   &   0.306   &  0.581   & 0.456  & 0.284  &  \underline{0.653}   &  0.550   &   \textbf{0.696}   \\
		AG      &   2.417   &   3.597   &   2.719   &  3.599   & 3.980  & 2.518  &  \underline{4.940}   &  \textbf{5.508}   &       4.688        \\
		SCD     &   1.396   &   1.798   &   1.619   &  \underline{1.805}   & 1.679  & 1.604  &  1.786   &  1.652   &   \textbf{1.834}   \\ \midrule
		\multicolumn{10}{c}{\textbf{Dataset: FLIR Image Fusion Dataset}}                           \\
		Metrics & FusionGAN & DenseFuse & ImageFuse & DeepFuse & TSIFVS & TVADMM & U2fusion & Nestfuse &      DIDFuse       \\ \midrule
		EN      &   7.017   &   7.213   &   6.992   &  7.213   & 7.152  & 6.797  &  7.279   &  \underline{7.316}   &   \textbf{7.392}   \\
		SD      &  34.383   &  37.315   &  32.579   &  37.351  & 35.889 & 28.071 &  40.985  &  \underline{43.640}  &  \textbf{51.968}   \\
		SF      &  11.507   &  15.496   &  14.519   &  15.471  & 18.794 & 14.044 &  \underline{19.217}  &  18.825  &  \textbf{22.101}   \\
		VIF     &   0.289   &   0.498   &   0.419   &  0.498   & 0.503  & 0.325  &  \underline{0.577}   &  0.563   &   \textbf{0.619}   \\
		AG      &   3.205   &   4.822   &   4.150   &  4.802   & 5.568  & 3.524  &  \underline{6.587}   &  6.444   &   \textbf{6.682}   \\
		SCD     &   1.182   &   \underline{1.716}   &   1.571   &  1.715   & 1.497  & 1.404  &  1.560   &  1.631   &   \textbf{1.816}   \\ \midrule
		\multicolumn{10}{c}{\textbf{Dataset: NIR Image Fusion Dataset, Country Scene}}                    \\
		Metrics & FusionGAN & DenseFuse & ImageFuse & DeepFuse & TSIFVS & TVADMM & U2fusion & Nestfuse &      DIDFuse       \\ \midrule
		EN      &   7.055   &   7.304   &   7.217   &  7.303   & 7.300  & 7.129  &  \underline{7.378}   &  7.361   &   \textbf{7.387}   \\
		SD      &  34.912   &  45.850   &  42.307   &  45.815  & 43.743 & 40.469 &  45.684  &  \underline{46.446}  &  \textbf{63.013}   \\
		SF      &  14.309   &  18.718   &  18.360   &  18.627  & 20.646 & 16.685 &  \underline{23.494}  &  20.579  &  \textbf{28.854}   \\
		VIF     &   0.424   &   0.677   &   0.613   &  0.676   & 0.688  & 0.530  &  \underline{0.789}   &  0.684   &   \textbf{1.016}   \\
		AG      &   4.564   &   6.228   &   5.920   &  6.178   & 6.823  & 5.319  &  \underline{8.682}   &  6.801   &   \textbf{9.660}   \\
		SCD     &   0.506   &   1.368   &   1.222   &  1.366   & 1.194  & 1.090  &  1.346   &  \underline{1.408}   &   \textbf{1.683}   \\ \midrule
		\multicolumn{10}{c}{\textbf{Dataset: NIR Image Fusion Dataset, Forest Scene}}                    \\
		Metrics & FusionGAN & DenseFuse & ImageFuse & DeepFuse & TSIFVS & TVADMM & U2fusion & Nestfuse &      DIDFuse       \\ \midrule
		EN      &   6.717   &   7.039   &   6.621   &  7.031   & 6.992  & 6.863  &  \underline{7.221}   &  7.154   &   \textbf{7.324}   \\
		SD      &  27.827   &  34.974   &  31.007   &  34.780  & 34.450 & 30.890 &  \underline{39.525}  &  36.468  &  \textbf{51.267}   \\
		SF      &  17.684   &  24.068   &  22.366   &  23.845  & 25.410 & 20.702 &  \underline{29.871}  &  24.894  &  \textbf{35.612}   \\
		VIF     &   0.466   &   0.790   &   0.669   &  0.787   & 0.798  & 0.645  &  \underline{0.930}   &  0.779   &   \textbf{1.234}   \\
		AG      &   6.254   &   8.975   &   8.016   &  8.864   & 9.207  & 7.658  &  \underline{12.333}  &  9.264   &  \textbf{13.166}   \\
		SCD     &   0.114   &   1.259   &   0.812   &  1.235   & 0.968  & 0.752  &  1.315   &  \underline{1.353}   &   \textbf{1.769}   \\ \midrule
		\multicolumn{10}{c}{\textbf{Dataset: NIR Image Fusion Dataset, Field Scene}}                     \\
		Metrics & FusionGAN & DenseFuse & ImageFuse & DeepFuse & TSIFVS & TVADMM & U2fusion & Nestfuse &      DIDFuse       \\ \midrule
		EN      &   6.723   &   7.000   &   6.559   &  7.014   & 6.974  & 6.776  &  \underline{7.167}   &  7.101   &   \textbf{7.287}   \\
		SD      &  30.627   &  41.280   &  34.551   &  41.572  & 38.577 & 35.626 &  43.345  &  \underline{43.953}  &  \textbf{59.332}   \\
		SF      &  14.837   &  18.704   &  16.832   &  18.845  & 19.443 & 15.717 &  \underline{23.105}  &  19.930  &  \textbf{27.081}   \\
		VIF     &   0.413   &   0.748   &   0.583   &  0.761   & 0.716  & 0.563  &  \underline{0.886}   &  0.754   &   \textbf{1.120}   \\
		AG      &   4.907   &   6.501   &   5.495   &  6.552   & 6.682  & 5.276  &  \underline{8.967}   &  6.776   &   \textbf{9.533}   \\
		SCD     &   0.359   &   1.361   &   0.958   &  1.412   & 1.099  & 0.975  &  1.426   &  \underline{1.475}   &   \textbf{1.787}   \\ \bottomrule
	\end{tabular}		
\end{table*}
	\begin{figure*}[h]
	\centering	\includegraphics[width=0.9\linewidth]{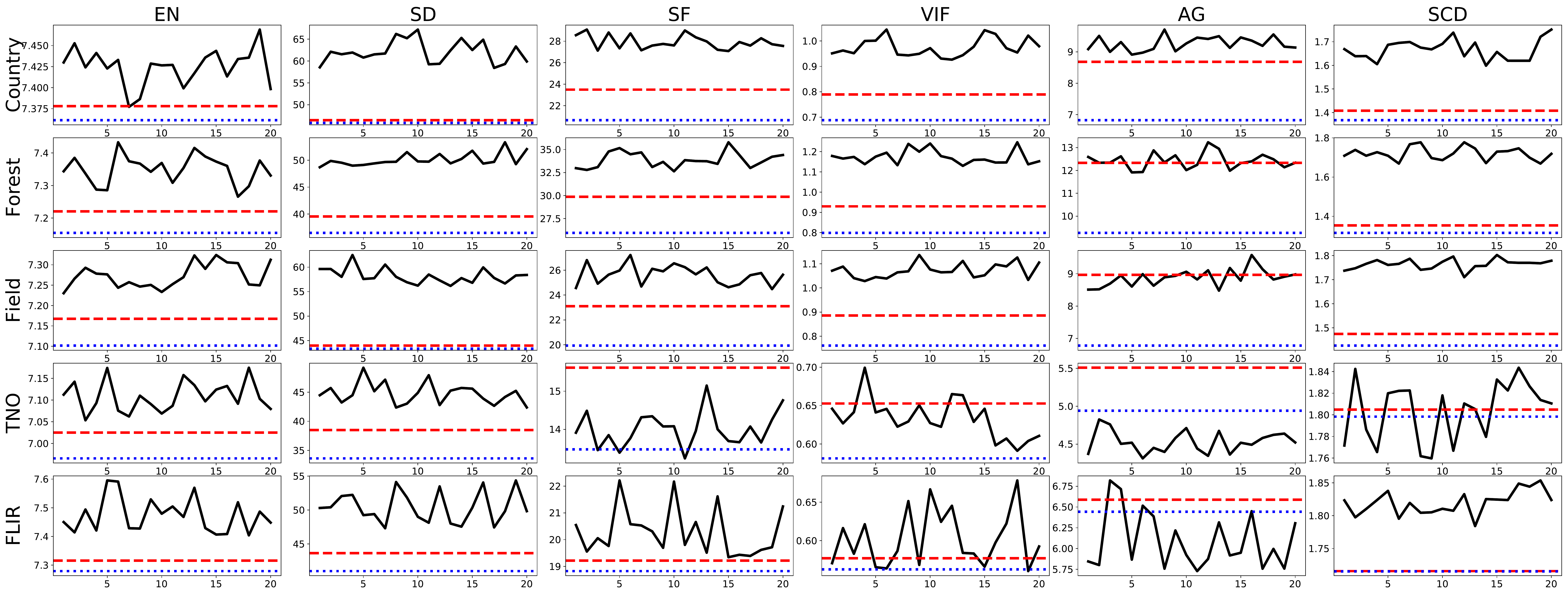}
	\caption{Test results of model reproducibility. From top to bottom: image fusion dataset NIR-Country, NIR-Forest, NIR-Field, TNO and FLIR. From left to right: the values of EN, SD, SF, VIF, AG and SCD.}
	\label{Robustness}
\end{figure*}
	
	\subsubsection{The role of skip connection}
	We do not change the structure or loss function of the original network, but only remove the skip connection module.
	
	In the above experiments, the evaluation metrics, the remaining tuning coefficients, training epoch, learning rate and other hyperparameters are consistent with our model. The FLIR, NIR-Mountain and NIR-Water datasets are still utilized as the training sets and ablation experiments are performed on the validation sets NIR-Urban and NIR-Street.
	
	Overall we count up the total number of larger values between the two compared models under each ablation experiment in Table~\ref{table_ablation}. Obviously, our model significantly contains larger values than the experimental group in each ablation experiment, which proves the effectiveness of each module and the rationality of its design.
	
	\subsection{Comparison with Other Models}
	In this subsection, we will compare our model with the other popular counterparts in the test datasets.
	
	\subsubsection{Qualitative comparison} 
	Figure~\ref{Figure} exhibits several representative fusion images generated by different models.
	Visual inspection shows that, in general, fusion results of our method obviously have higher contrast, more details, and clearer target presentation.
	In detail, if the fusion images containing people~(the first, third, and fourth columns), other methods have problems such as weak high-lighted objects, poor contrast and less prominent contour of targets and backgrounds. In the first column, the person with an umbrella is more salient. The edge of houses in the third column is easier to distinguish. The radiation information of individuals in the fourth column is better preserved, and clouds in the background are also particularly clear.
	Similarly, if the images are natural landscapes~(the second, fifth, and sixth columns), other methods have blurred boundaries, poor color contrast, and insufficient sharpness.
	Conversely, our method can obtain fused images with brighter targets, sharper edge contours and retaining richer detailed information, such as the brighter bunker in the second column, the clearer edges of mountains and clouds in the fifth column, and the leaves with abundant texture in the sixth column, etc.
	
	\subsubsection{Quantitative comparison}
	Subsequently, quantitative comparison results on test sets are listed in Table \ref{table}. It is found that our model is the best performer on all datasets in terms of almost all metrics. As for competitors, they may perform well on one dataset in terms of part of metrics. This result demonstrates that images fused by our model are with enriched textures, high contrast and satisfy the human visual system, successfully retained most of the detailed information of the source images.
	\subsection{Experiments on Reproducibility}

	As is known, deep learning methods are often criticized for instability. Therefore, we test the reproducibility of DIDFuse in the last experiment. We repeatedly train the network 20 times and quantitatively compare the 20 parallel results. As shown in Figure \ref{Robustness}, the black solid curves report six metrics over 20 experiments. The red dashed line and blue dotted line represent the greatest and the second greatest values in the compared methods, respectively.
	Similar to the above results, our method can basically keep the first place all the time, indicating that DIDFuse can generate high-quality fused images steadily.

	\section{Conclusion}\label{sec:5}
	Aiming at solving the IVIF task, a new dual-stream AE network is constructed in which the encoder is responsible for two-scale image decomposition and the decoder is employed for original image reconstruction.
	In the training phase, the encoder is trained to output base and detail feature maps, then the decoder reconstructs original images.
	In the test phase, we set a fusion layer between the encoder and decoder to fuse base and detail feature maps through an attention mechanism based fusion strategy. Finally, the fused image can be acquired through the decoder.
	We test our model on TNO, FLIR, and NIR datasets. Qualitative and quantitative results show that our model outperforms other SOTA methods, since our model can steadily obtain fusion images containing highlighted targets and rich details.
	
	In the future, we will consider applying this model to other image fusion tasks, and explore the application of this complementary information extraction model in other image processing fields.

	\ifCLASSOPTIONcaptionsoff
	\newpage
	\fi
	
	\bibliographystyle{IEEEtran}
	\bibliography{xbib}
	
	\begin{IEEEbiography}[{\includegraphics[width=1in,height=1.25in,clip,keepaspectratio]{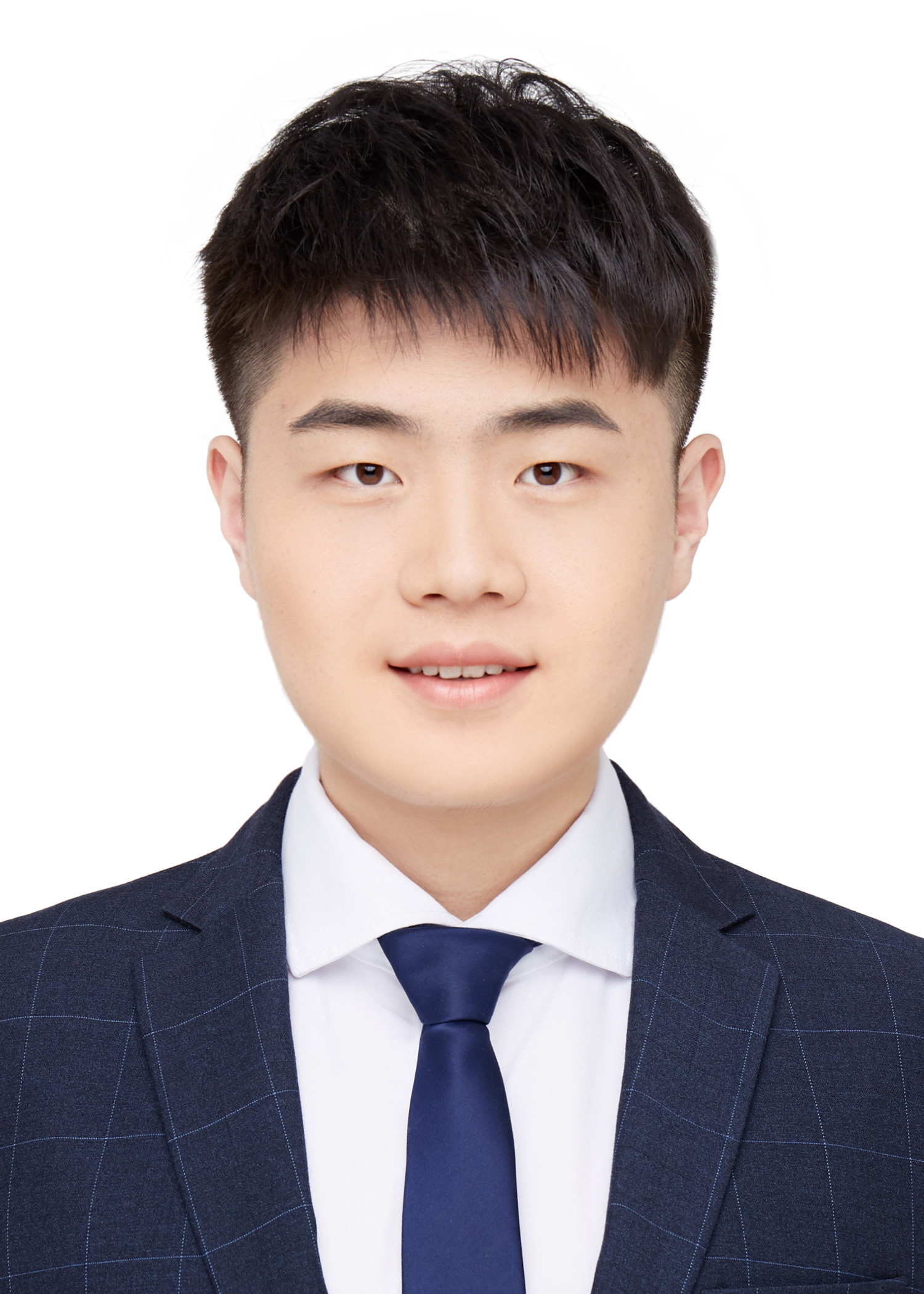}}]{Zixiang Zhao} is currently pursuing the Ph.D. degree in statistics with the School of Mathematics and Statistics, Xi’an Jiaotong University, Xi’an, China. His research interests include computer vision, deep learning and low-level vision.\end{IEEEbiography}
	\begin{IEEEbiography}[{\includegraphics[width=1in,height=1.25in,clip,keepaspectratio]{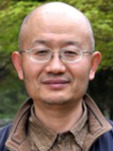}}]{Jiangshe Zhang} received the B.S., M.S., and Ph.D. degrees in computational mathematics from Xi’an Jiaotong University, Xi’an, China, in 1984, 1987, and 1993, respectively.
		
		He is currently the Director of the Institute of Machine Learning and Statistical Decision Making, Xi’an Jiaotong University, where he is also a Professor with the Department of Statistics. He is also the Vice-President of the Xi’an International Academy for Mathematics and Mathematical Technology, Xi’an. He has authored and co-authored one monograph and over 100 journal papers. His current research interests include statistical computing, deep learning, cognitive representation, and statistical decision making.
		
		Dr. Zhang received the National Natural Science Award of China (Third Place) in 2007 and the First Prize in Natural Science from the Ministry of Education of China in 2007. He served as the President of the Shaanxi Mathematical Society and the Executive Director of the China Mathematical Society.
	\end{IEEEbiography}
	\begin{IEEEbiography}[{\includegraphics[width=1in,height=1.25in,clip,keepaspectratio]{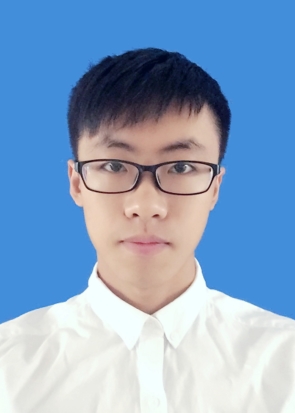}}]{Shuang Xu} is currently pursing the Ph.D. degree in statistics with the School of Mathematics and Statistics, Xi'an Jiaotong University, Xi'an, China. His current research interests include Bayesian statistics, deep learning and complex network.
	\end{IEEEbiography}
	\begin{IEEEbiography}[{\includegraphics[width=1in,height=1.25in,clip,keepaspectratio]{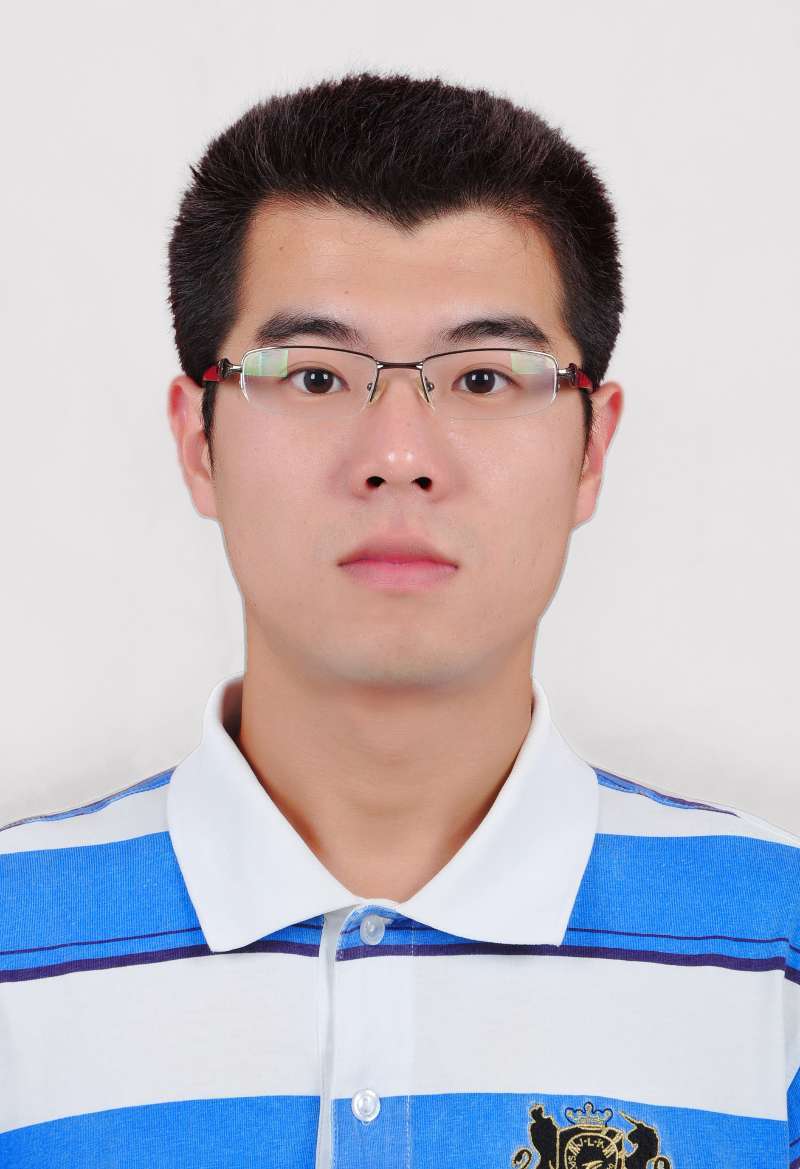}}]{Kai Sun}
		received his Ph.D degree in statistics from Xi'an Jiaotong University, Xi'an, China, in 2020. He is currently a postdoctoral research fellow with Xi’an Jiaotong University,
		Xi’an, China. His current research interests include deep learning, pattern recognition, and image processing.
	\end{IEEEbiography}
	\begin{IEEEbiography}[{\includegraphics[width=1in,height=1.25in,clip,keepaspectratio]{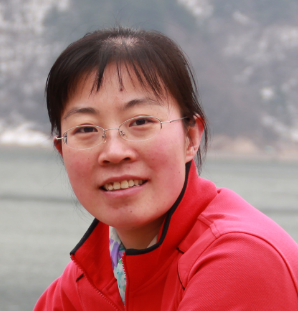}}]{Chunxia Zhang} received her Ph.D degree in Applied Mathematics from Xi'an Jiaotong University, Xi'an,  China, in 2010.
	
	Currently, she is a Professor in School of Mathematics and Statistics at Xi'an Jiaotong University. She has authored and coauthored about 30 journal papers on ensemble learning techniques, nonparametric regression and etc. Her main interests are in the area of ensemble learning, variable selection and deep learning.
	\end{IEEEbiography}
	\begin{IEEEbiography}[{\includegraphics[width=1in,height=1.25in,clip,keepaspectratio]{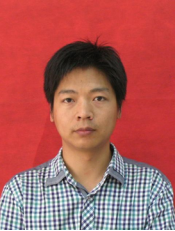}}]{Junmin Liu} (M'13)
	received the M.S. degree in computational mathematics from Ningxia University, Yinchuan, China, in 2009, and the Ph.D. degree in applied mathematics from Xi’an Jiaotong University, Xi’an, China, in 2013. 
	
	He is currently an Associate Professor with the School of Mathematics and Statistics, Xi’an Jiaotong University. His current research interests include hyperspectral unmixing, remotely sensed image
	fusion, and deep learning.	\end{IEEEbiography}
	
\end{document}